\documentclass{article}
\usepackage[utf8]{inputenc}
\usepackage{authblk}
\usepackage{setspace}
\usepackage[margin=1.25in]{geometry}
\usepackage{graphicx}
\graphicspath{ {./figures/} }
\usepackage{subcaption}
\usepackage{amsmath}
\usepackage{lineno}

\usepackage[style=nejm, 
citestyle=numeric-comp,
sorting=none]{biblatex}
\title{Synthesis of electron microbunching rotation for generating isolated attosecond soft x-ray free-electron laser pulses}

\author[1]{Hao Sun}
\author[1*]{Xiaofan Wang}
\author[1]{Li Zeng}
\author[2*]{Weiqing Zhang}

\affil[1]{Institute of Advanced Science Facilities, Shenzhen 518107, China}
\affil[2]{Dalian Institute of Chemical Physics, Chinese Academy of Sciences, Dalian 116023, China}
\affil[*]{Address correspondence to: wangxf@mail.iasf.ac.cn and weiqingzhang@dicp.ac.cn}

\date{}

\begin{document}

\maketitle

\begin{abstract}
Attosecond X-ray pulses play a crucial role in the study of ultrafast phenomena involving inner and valence electrons. Especially isolated attosecond pulses with high photon energy and high peak power are of great significance in single-shot imaging in the soft X-ray region, life sciences, and attosecond pump-probe experiments. In modern accelerators, laser manipulation of electrons can be used to tailor the ultrafast properties of free-electron laser pulses. In this paper, we propose a novel laser manipulation technique that makes use of two many-cycle laser beams with mutual delays and tilted wavefronts to synthesize microbunching rotation on the scale of infrared laser wavelengths within the electron bunch. This synthesis microbunching rotation ultimately leads to an enhanced current contrast ratio between the main peak and the surrounding satellite peaks within the bunch. By properly accounting for the longitudinal space charge fields within the undulator, a tapered undulator can further suppress the side peaks in the radiation pulse and enable the selection of an isolated, hundred-attosecond, GW-level soft X-ray pulse.

\end{abstract}


\section{Introduction}

Attosecond $\left(1\right.$ as $\left.=10^{-18}~\mathrm{s}\right)$ soft X-ray sources hold the potential to precisely measure and control the movement of bound electrons in matter, which is expected to catalyze groundbreaking discoveries in the realm of ultrafast science \cite{attosecond1,attosecond2}. The plenitude of atomic resonances and effective photoabsorption within the soft X-ray spectral range has sparked significant interest in the generation of intense attosecond soft X-ray pulses, particularly within the atomic, molecular, and optical scientific community \cite{ESASE_6}. Historically, high harmonic generation (HHG) sources have been the cutting-edge approach for generating attosecond pulses \cite{HHG1,HHG2,HHG3,HHG4,HHG5}.
The entire HHG process takes place within the driving laser optical
cycle and repeats every half-cycle, leading to a train of high
harmonic bursts each with a sub-femtosecond pulse duration. 
With restricting these bursts to within one-half cycle of the laser field, a single attosecond pulse may be isolated, which is crucial for studying electron dynamics in pump-probe experiments \cite{HHG1}. In order to obtain an isolated attosecond pulse, different gating techniques have progressively emerged. These include spectral selection of half-cycle cutoffs \cite{Attosecond_control} through amplitude gating \cite{science.1157846} and ionization gating \cite{HHG2}, temporal gating such as polarization gating \cite{science.1132838} and double optical gating \cite{PhysRevLett.100.103906}, and spatiotemporal gating using the attosecond lighthouse effect \cite{attoseond_light_house}. However, their pulse energy output does not exhibit a favorable scaling relationship with the photon energy \cite{HHG5}. Consequently, the energy of attosecond pulses in the soft X-ray region is typically limited, which restricts their usage in nonlinear experiments, attosecond pump-probe spectroscopy \cite{science.adn6059}, and single-shot ultrafast imaging within the soft X-ray domain.

X-ray free-electron lasers (FELs) \cite{doi:10.1126/science.1055718,McNeil2010,Seddon_2017} offer a compelling solution to overcome the intensity and photon energy limitations of the HHG-based radiation source, offering a distinct pathway for generating ultrashort radiation pulses with exceptional brightness. Recently, attosecond science has been rapidly advancing within the FEL community, thanks to various proposed schemes such as the chirp-taper scheme \cite{chirp_taper1,chirp_taper2}, enhanced self-amplified spontaneous emission scheme (ESASE) \cite{ESASE_2}, and seeded FEL schemes \cite{Nature_atto_seededFEL,PhysRevSTAB.12.060701,Xiao}. The laser-assisted ESASE technique leverages few-cycle lasers to induce a time-dependent energy modulation within the bunch \cite{ESASE_2}. By incorporating a dispersive section, this technique enhances the local peak current, enabling the generation of ultrashort attosecond pulses of hundreds of attoseconds. However, the presence of satellite current spikes in the ESASE technique hampers the generation of isolated attosecond pulses. While the power of the side peaks may not be as high as that of the main peak, they still reside within the same order of magnitude. In practical applications, the energy of the side peaks is sufficiently potent to excite the sample, thus introducing interference to the results of ultrafast measurement experiments. To suppress the generation of satellite spikes in ESASE, Ding extended the attosecond, two-color ESASE scheme proposed by Zholents to a long optical cycle system using a second detuned laser and a tapered undulator \cite{ESASE_3}. In order to generate isolated attosecond pulses in other FEL schemes, several methods have been proposed and validated. In the SLAC National Accelerator Laboratory, the X-ray laser-enhanced attosecond pulse generation (XLEAP) project demonstrated the generation of isolated soft X-ray attosecond pulses with gigawatt peak power \cite{ESASE_5,TW}.
Tanaka proposed a novel scheme to generate an isolated monocycle X-ray pulse in free-electron lasers, which is based on coherent emission from a chirped microbunch passing through a strongly tapered undulator \cite{PhysRevLett.114.044801}.
Remarkably, this approach has recently been partially validated on a storage ring \cite{PhysRevLett.131.145001}.
Recently, a conceptual design for a terawatt-level, isolated at the sub-attosecond pulse of 248 keV photon energy, is proposed through a simulation study \cite{APL10.1063/5.0067074}.

In modern accelerators, some promising techniques that rely on lasers to manipulate and rearrange the electron distribution have been proposed in order to tailor the properties of the radiation \cite{RevModPhys.86.897,Zholents_2008,doi:10.1126/science.287.5461.2237}. Previous studies have shown that the interaction between a tilted laser and an electron beam in an undulator can achieve three-dimensional manipulation of the electron beam \cite{PhysRevSTAB.17.070701,PhysRevAccelBeams.22.070701,Wang_2020,LU2022105849,Feng:15}. Noncollinear optical gating is a technique that utilizes two identical laser pulses with a time delay and noncollinear overlap to create a driving field with rotating wavefronts \cite{Heyl_2014,Kennedy_2022}. This approach is employed to generate angularly streaked attosecond pulse trains, serving as an effective gating mechanism for the generation of isolated attosecond pulses. This method has been experimentally validated and applied in HHG \cite{Louisy:15}. In this paper, we propose a novel laser manipulation technique for generating isolated attosecond pulses by synthesizing electron microbunching rotation with two tilted-wavefront many-cycle lasers. In the proposed scheme, two many-cycle laser beams with mutual delays and tilted wavefronts interact with an electron beam in a wiggler, leading to a multidimensional modulation. Subsequently, the electron beam undergoes density modulation after passing through a chicane, resulting in microbunching at the wavelength scale of the external laser. At this point, the microbunching within the electron beam experiences wavefront rotation, with the overlapping region of the two laser beams exhibiting the highest current distribution, while the current in other regions decreases due to wavefront rotation. Furthermore, a tapered undulator is employed to mitigate the degradation effect of the longitudinal space charge effect in the main current spike while also suppressing the FEL gain of all side current spikes. The remainder of the paper is organized as follows: In Sec.~\ref{sec2}, we provide an overview of the principles behind the proposed scheme. In Sec.~\ref{sec3}, we present simulation results to demonstrate our scheme.  Finally, we provide discussions and a brief summary of this work in Sec.~\ref{sec4}.

\section{Principle}\label{sec2}
\begin{figure*} 
\includegraphics[width=15cm]{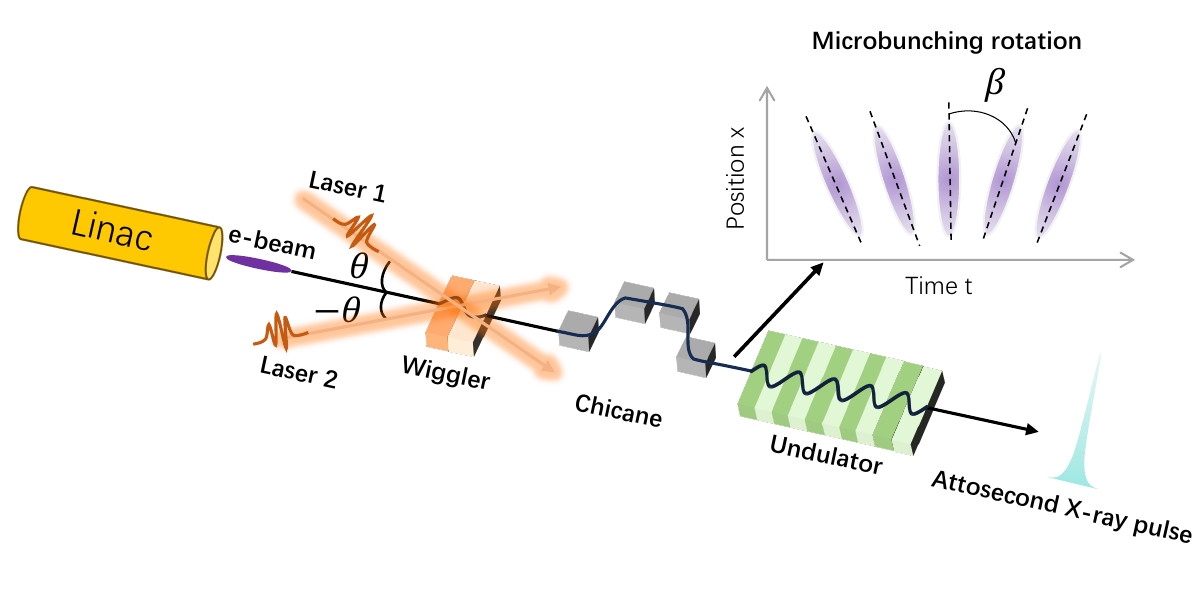}
\caption{ Schematic of the proposed scheme for generating attosecond FEL pulses.}
\label{figure1}
\end{figure*}

The schematic of the proposed scheme, as sketched in  Fig.~\ref{figure1}, consists of two identical laser pulses, separated by a time $\Delta t$ incident on a wiggler with an angle $2\theta$ between them. The introduction of a temporal delay between two incident laser pulses leads to an amplitude ratio that changes rapidly from one half-cycle to the next, as shown in Fig. 2(a), defining a unique orientation angle of the corresponding wave fronts as a function of time. The consequence is an ultrafast wavefront rotation, as shown in Fig. 2(b). Here we assume an initial Gaussian beam energy
distribution with an average energy $\gamma_0 m c^2$, and we introduce the
variable $p=\left(\gamma-\gamma_0\right) / \sigma_\gamma$ to represent the dimensionless energy deviation of a particle, where $\sigma_\gamma$ is the initial beam energy spread. Thus, the initial longitudinal phase space distribution should be written as $f_0(p)=N_0 \exp \left(-p^2 / 2\right) / \sqrt{2 \pi}$. The initial transverse electron beam distribution can be written as $g_0(X)=N_0 \exp \left(-X^2 / 2\right) / \sqrt{2 \pi}$, and  $X=\left(x-x_0\right) / \sigma_x$ is the dimensionless horizontal position of a particle, where $\sigma_x$ is the initial beam size. After passing through the wiggler, the dimensionless energy deviation of the electron beam becomes \cite{PhysRevSTAB.17.070701}
\begin{equation}
p^{\prime}=p+A_1(t) \sin \left(k_s z+k_s \tan (\theta) x\right)+A_2(t) \sin \left(k_s z+k_s \tan (-\theta) x\right) ,
\end{equation}
where $k_s$ is the wave number in the longitudinal direction, $k_s \tan \theta$ is the wave number in the horizontal direction due to the relative wavefront-tilt, and $A_1(t)$, $A_2(t)$ is the energy modulation induced by laser 1 and laser 2, respectively. Since the two lasers are identical, the maximum value of  $A_1(t)$ is the same as the maximum value of $A_2(t)$. When x is infinitely close to zero, the above equation can be simplified to:
\begin{equation}
p^{\prime}=p+A(t) \sin \left(k_s z+k_s \beta(t) x\right),
\end{equation}
where $\beta(t)=\tan (\theta) \frac{1-\xi(t)}{1+\xi(t)}$, $\xi(t)=A_2(t) / A_1(t)$, $A(t)=\Delta \gamma(t) / \sigma_\gamma$ is the total energy modulation amplitude, and $\Delta \gamma(t)$ is the total time-dependent energy modulation depth induced by the two incident laser.
The two-dimensional distribution function of the electron beam after the wiggler can be presented as:
\begin{equation}
h_1(\zeta, p, X)=\frac{N_0}{\sqrt{4 \pi^2}} \exp \left\{-\frac{1}{2}\left[p-A(t) \sin \left(\zeta-k_s \beta(t)\sigma_x X\right)\right]^2\right\}\exp \left(-X^2 / 2\right),
\end{equation}
where $\zeta=k_s z$ is the phase of the electron beam. After passing through the chicane with the dispersive strength of $R_{56}$, two-dimensional distribution function of the electron beam evolves to
\begin{equation}
h_2(\zeta, p, X)=\frac{N_0}{\sqrt{4 \pi^2}} \exp \left\{-\frac{1}{2}\left[p-A(t) \sin \left(\zeta-k_s \beta(t)\sigma_x X-Bp\right)\right]^2\right\}\exp \left(-X^2 / 2\right),
\end{equation}
where $B=R_{56} k_s \sigma_\gamma / \gamma$ is the dimensionless strength of the chicane. Integration of Eq. (6) over p and x gives the beam density $N$ as a function of $\zeta$,

\begin{equation}
N(\zeta)=\int_{-\infty}^{\infty} d x \int_{-\infty}^{\infty} d p h_2(\zeta, p, X).
\end{equation}
The current distribution of the electron beam can be represented as:

\begin{equation}
\begin{aligned}
I( \zeta)=\frac{N(\zeta)}{N_0}= & 1+2 \sum_{n=1}^{\infty} J_n\left[n A(t) B)\right] 
\times \exp \left[-\frac{1}{2} n^2 B^2\right] 
 \times \exp \left[-\frac{1}{2} {n^2 {k_s^2 (\beta(t) \sigma_x)^2}}\right] \cos (n \zeta).
\end{aligned}
\end{equation}
It can be observed that the current distribution depends on the rotation angle $\beta(t)$. It is evident that the smaller the product $\beta(t)\sigma_x$, the higher the peak current, and the current reaches its maximum value when $\beta(t)$ equals zero. Therefore, this approach allows for controlling the current distribution by manipulating the rotation angle $\beta(t)$, thus generating isolated attosecond pulses with a higher signal-to-noise ratio.

\begin{figure}[h]
    \centering
    \includegraphics[width=0.8\textwidth]{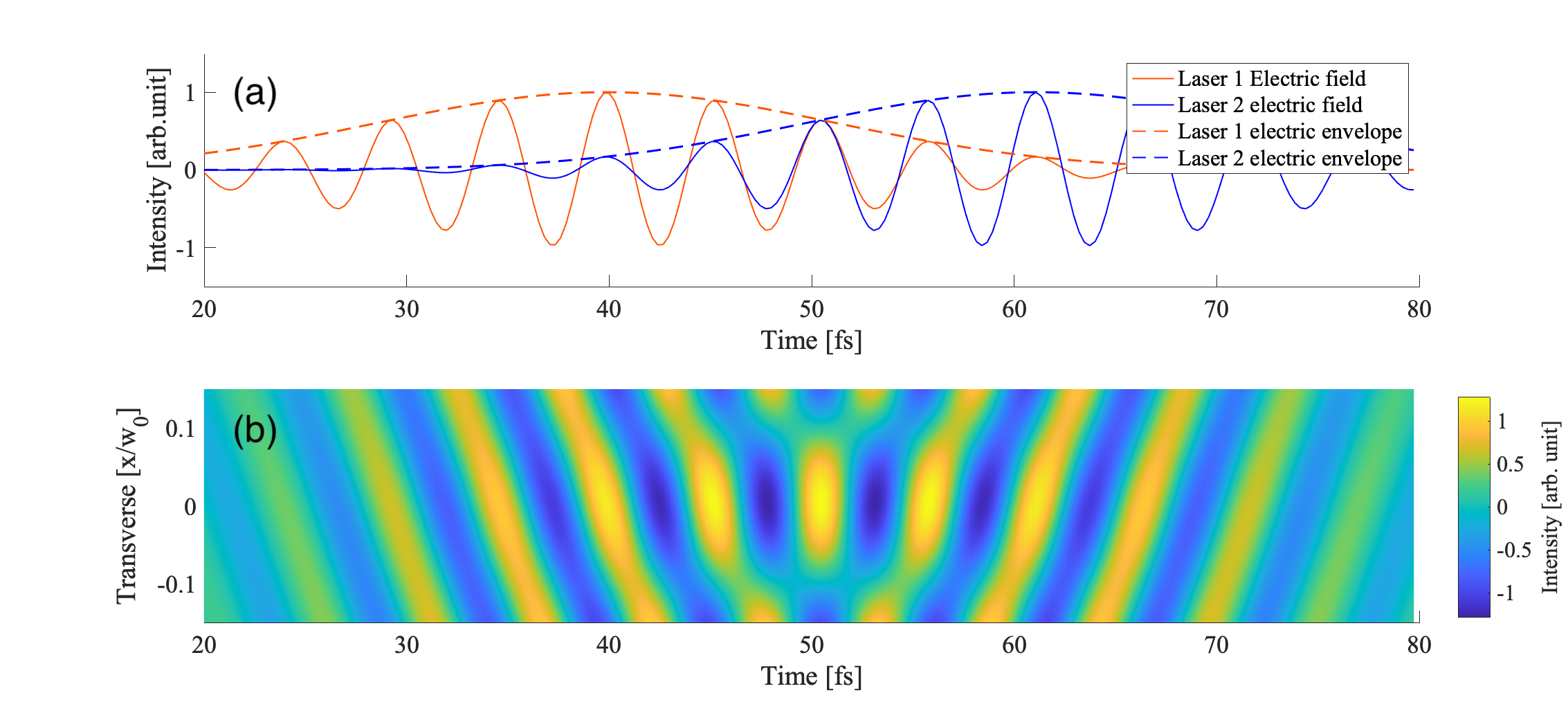}
\caption{(a) Electric field of two lasers with a mutual delay. (b) ${Real} (E)$ of the total field $E$ after superposition of two few-cycle laser pulses. (${w}_0$ is the radius of the waist radius.)}
    \label{figure2}
\end{figure} 

\section{Results} \label{sec3}
To illustrate the physical mechanism of the proposed scheme, three-dimensional simulations were performed with realistic parameters of a typical soft X-ray free-electron laser facility, as listed in Table \ref{table1}. The electron beam extracted from a linear accelerator was characterized by the following parameters: an energy of 2.53 GeV, a peak current of 0.9 kA, rms normalized emittance of 0.45~mm$\cdot$mrad, and an uncorrelated energy spread of 250 keV. Table \ref{table1} also summarizes other parameters for the external lasers, wiggler, and undulator employed in this study. In this simulation, the interaction between the tilt incident laser and the electron beam in a wiggler was performed by a three-dimensional algorithm based on the fundamental electrodynamic theory when considering the appearance of electric and magnet fields of a wavefront-tilted seed laser beam \cite{PhysRevAccelBeams.22.070701,photonics7040117}. The three-dimensional time-dependent FEL process in an undulator was simulated by Genesis 1.3 \cite{reiche1999genesis}.
\begin{table}[h]
    \caption{Parameters in the simulations.
}    
    \centering
    \begin{tabular}{ccc}
            \hline
            Parameter & Value \\  
            \hline
            \textbf{Electron beam} &  \\ 
            \hline
            Energy  & 2.53 GeV  \\
            Peak current & 900 A \\
            Emittance & 0.45 mm$\cdot$mrad  \\
            rms energy spread  & 250 keV \\
            \hline
            \textbf{Laser} &  \\ 
            \hline
            Laser-1/Laser-2 wavelength & 1600/1600 nm \\
            Laser-1/Laser-2 FWHM duration & 18.8/18.8 fs \\
            Laser-1/Laser-2 peak power & $24/24 \mathrm{~GW}$\\
            Laser-1/Laser-2 incidence angle $\theta$ & 4/-4 mrad \\
            Laser-1/Laser-2 time delay &21.2 fs\\
            \hline
            \textbf{Wiggler} &  \\ 
            \hline
            Wiggler period $\lambda_{wig}$/number & 15 cm/2 \\
            Wiggler magnetic field $B_{\theta}$ & 1.15 T\\
            \hline
            \textbf{Chicane} &  \\ 
            \hline
            $R_{56}$ & $\sim 0.8 \mathrm{~mm} $\\
            
            \hline
            \textbf{FEL} &  \\ 
            \hline
             Main undulator period/number & 3 cm/587 \\
             Average beta function& 8 m\\
             Photon energy &620 eV\\
            \hline
             \end{tabular}
    \label{table1}
\end{table}
\subsection{Microbunching rotation}
When a laser is incident on the wiggler at an oblique angle $ \theta$, the off-axis resonance condition is

\begin{equation}
\lambda_L=\frac{\lambda_{wig}}{2 \gamma^2}\left(1+\frac{{K_\theta}^2}{2}+\gamma^2 \theta^2\right),
\end{equation}
where $\lambda_L$ is the laser wavelength, and $\lambda_{wig}$ is the period length of the wiggler. The magnetic field intensity of the wiggler can be calculated using the following formula:
\begin{equation}
B_{\theta}=\frac{2  K_\theta  \pi m c}{\lambda_{wig} e},
\end{equation}
where $e$ is the single electron charge, and $c$ is the light speed.
The calculation assumes $\lambda_{wig}$ remains unchanged when we adjust the incident angle $\theta$. This leads to a limited region of $\theta$ to satisfy the resonant condition, as shown in Fig. 3(a). The angular deviation between the laser and electron beams limits their effective interaction region. For instance, when the incident angle $\theta=4 ~\mathrm{mrad}$ and the wiggler length is $L_{wig}=0.3 \mathrm{~m}$, the separation between the center of the electron beam and the center of the laser beam at the entrance and exit of the undulator is $\frac{L_{wig}}{2} \theta= 0.6~\mathrm{~mm}$. Therefore, in order to enhance the effective interaction between the electron beam and the laser at larger incidence angles, we have chosen a relatively large laser radius of 1 mm. By employing the aforementioned parameters, simulations were conducted to investigate the relationship between the energy modulation depth and the peak power of the laser, as well as the incidence angle $\theta$. It is evident that a significant increase in laser power is required to achieve the same depth of energy modulation at a larger laser incidence angle.
\begin{figure}[h]
    \centering
    \includegraphics[width=0.8\textwidth]{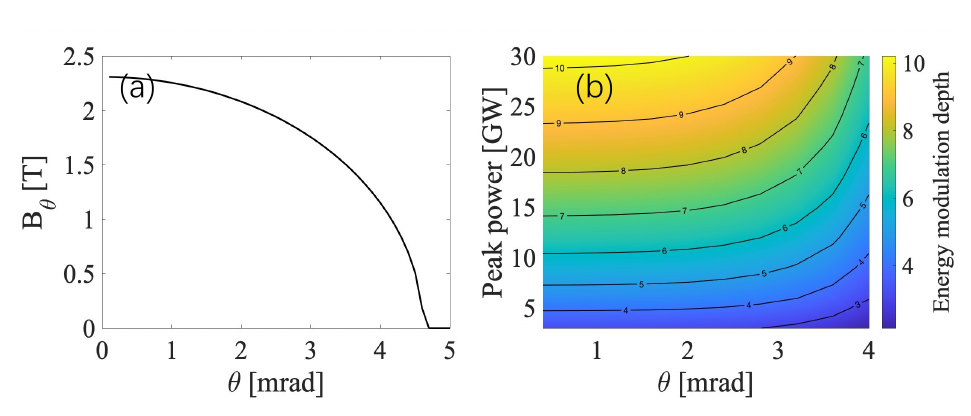}
    \caption{(a) Relation between the magnetic field of the wiggler and the laser incidence angle. (b) Depth of energy modulation (Ratio of laser-introduced energy modulation to slice energy spread) as a function of the laser peak power and the laser incident angle.}
    \label{figure3}
\end{figure}
\begin{figure}[h]
    \centering
    \includegraphics[width=0.8\textwidth]{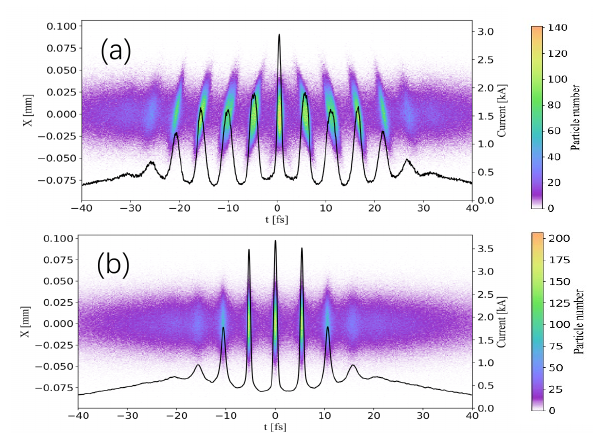}
    \caption{Spatial distribution of the electron beam and the corresponding current distribution in the proposed scheme (a) and in the standard ESASE scheme (b) where only one many-cycle laser without any wavefront tilt is employed.}
    \label{figure4}
\end{figure}

Figure 4(a) illustrates the transverse beam distribution and corresponding current distribution of the electron beam after passing through the wiggler and chicane based on the parameters listed in Table~\ref{table1}. The transverse beam distribution of the electron beam exhibits a fascinating wavefront rotation pattern. The central region of the beam shows a non-tilted transverse distribution, while the outer regions display a tilted distribution. Consequently, the central region experiences a higher current with a narrower width, while the side regions demonstrate a lower current with a wider width. In contrast, when the electron beam is modulated by one many-cycle laser without any wavefront tilt, the transverse distribution of the electron beam after passing through the chicane is shown in Fig. 4(b). It can be observed that the electron beam does not exhibit a tilted transverse distribution, and the current distribution in the center region is roughly similar to that of the sides.

\subsection{Longitudinal space charge effect}

After passing through the chicane, the energy modulation is transformed into a density modulation, resulting in a peak current of 3 kA in the central spike, as illustrated in Fig. 4(a). Since only a small portion of the bunch charge is concentrated in the spike region, the longitudinal space charge (LSC) effect is not negligible \cite{ESASE_3}. Due to the wiggling motion in the undulator, the longitudinal space charge field can be approximated as the result in free space by substituting $\gamma$ with $\bar{\gamma}_z$, where $\bar{\gamma}_z=\gamma / \sqrt{1+K^2 / 2}$, and $K$ represents the undulator parameter. A simplified expression can be utilized to estimate the longitudinal space charge field \cite{ESASE_3}:

\begin{equation}
E_z \approx-\frac{Z_0 I^{\prime}(s)}{4 \pi \bar{\gamma}_z^2}\left(2 \ln \frac{\bar{\gamma}_z \sigma_z}{r_b}+1-\frac{r^2}{r_b^2}\right),
\end{equation}
where $r=\sqrt{x^2+y^2}$, $Z_0=377~\Omega$, $I^{\prime}(s)=d I / d s$ is the derivative of the electron current profile with respect to the longitudinal bunch coordinate $s$, $\sigma_z$ denotes the rms bunch length of the current spike, and $r_b$ is the beam radius of a uniform transverse
distribution. It is worth noting that for a tilted microbunching, if $\sigma_x \theta>1 / k_s$, the 1D LSC calculations can underestimate the LSC effect by a large factor \cite{PhysRevSTAB.13.020703}. However, the 1D LSC calculations remain appropriate since $\sigma_x \theta<1 / k_s$ is satisfied in the simulations. Here, we take $\bar{\gamma}_z \sim 3389$, $r_b \approx 2 \sigma_x \sim  50~\mu \mathrm{m}$, and $\sigma_z\sim 130 \mathrm{~nm}$ for the central spike. With these parameters, $E_z$ exhibits very weak dependency on the transverse position of the electrons within the beam. Hence, we drop the r-dependent term when calculating the LSC field. Fig. 5 shows the LSC field in the undulator. The longitudinal space charge effect can significantly degrade the FEL performance, which we are taking into account in the following simulations.

\begin{figure}[h]
    \centering
    \includegraphics[width=0.8\textwidth]{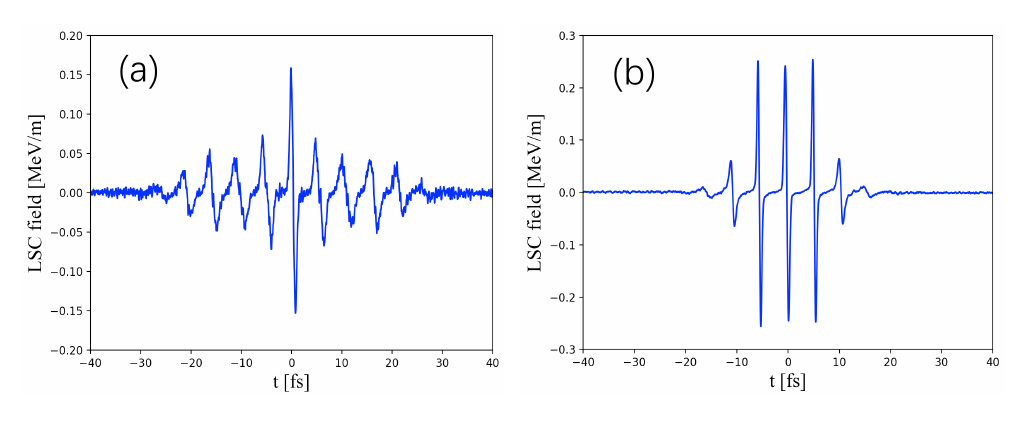}
    \caption{Longitudinal space charge field in the undulator considered in the proposed scheme (a) and the standard ESASE scheme (b).}
    \label{figure5}
\end{figure}

\subsection{Attosecond FEL performances}

The research \cite{chirp_taper1} highlighted that a tapered undulator can counterbalance the decline in FEL gain caused by a linearly chirped electron beam entering the undulator. When employing the ESASE scheme with high-current spikes, the LSC field within the FEL undulator induces an approximately linear energy chirp within these spikes, which also grows proportionally with the undulator distance. As the radiation wave advances towards the front of the current spike, LSC leads to an energy increase in the front portion of the spike, thereby offsetting the resonant condition. By adjusting the undulator parameter in a tapered undulator, we can largely compensate for the energy variation and maintain the resonant condition for the interacting electrons. Nevertheless, since the strength of the LSC fields relies on the derivative of the current, the appropriate taper strength is greater for the central peak compared to the side peaks. In our approach, the current for the central peak is approximately twice as large as that for the side peaks, as shown in Fig. 4(a). By aligning the taper to match the strongest chirp of the central peak, we can sustain the resonant condition in this region while simultaneously suppressing the FEL process in other parts of the bunch. Furthermore, the higher current in the central peak results in a greater FEL gain compared to the side peaks. Consequently, the combination of the proposed current modulation and undulator taper yields a favorable contrast ratio between the central and side X-ray spikes. 

When the energy chirp is mainly accumulated in the undulator and the bunch is sufficiently short, the LSC effect can be compensated by a quadratic taper \cite{chirp_taper7}:
\begin{equation}
\frac{d^2 K}{d z^2}=-\frac{\left(1+K_0^2 / 2\right)^2}{K_0} \frac{1}{\gamma_0^3} \frac{d^2 \gamma}{cd t d z}.
\end{equation}
We utilize Genesis 1.3 to determine the optimal undulator taper that maximizes the contrast ratio between the central and side spikes in the radiation pulse. The main undulator is a variable-gap undulator that spanning length of 17.6 meters, with a reverse quadratic taper of $d^2 K / d z^2=8 \times 10^{-5} \mathrm{~m}^{-2}$. The stability of SASE FEL is affected by inherent noise jitter in the electron beam, and in order to more accurately assess the performance of the scheme, 30-shot simulations were performed with different noise realizations.
Figure 6 summarizes the FEL simulation results at wavelength of 2 nm. The multi-shot power and spectral distributions for the proposed scheme are presented in Fig. 6(a) and Fig. 6(b), respectively.
The contrast of the isolated spike is over $98 \%$ in the proposed scheme. The statistical distribution of peak power is $1.1 \pm 0.6~\mathrm{GW}$, and the statistical distribution of the FWHM length is $804.0 \pm 184.5~\mathrm{as}$. 
It is worth emphasising that the initial peak current of the electron beam in this simulation is only 0.9 kA, so after processes such as ESASE, the peak current of the beam goes to roughly 3 kA, which results in a peak power of up to several gigawatts of FEL radiation. At higher beam powers, the synthesis approach is fully feasible to generate peak power of tens of gigawatts or even higher. In order to compare the contrast of attosecond FEL pulses  between the proposed scheme and the standard ESASE scheme under many-cycle laser conditions, FEL simulations were carried out using the standard ESASE beam distribution shown in Fig. 4(b). In the standard ESASE scheme, only one many-cycle laser without any wavefront tilt is used for the modulation process. The multi-shot power and spectral distributions are displayed in Fig. 6(c) and Fig. 6(d), respectively. After considering the LSC effect shown in Fig. 5(b) as well as the reverse taper compensation in the simulation, the peak power at the undulator exit, of the same length as in the proposed scheme, measures $0.6 \pm 0.3~\mathrm{GW}$. However, it can be observed that there are strong satellite peaks on either side of the main spike. Furthermore, we also considered the impact of the relative time jitter between the two external laser beams on the scheme.
Figure 7(a) presents the FEL performances under different relative time jitters between the two lasers. When the peak-to-peak relative time jitter between the two laser beams is within 2 fs, the contrast of isolated attosecond pulses is $90 \%$. With the current state-of-the-art laser technology, the jitter requirements for two lasers from the same source can be achieved. Finally, we also consider the contrast of isolated attosecond pulses under different laser incident angles. Figure 7(b) presents the FEL performances under different laser incident angles. When the laser incident angle is reduced to $3 ~\mathrm{mrad}$, the contrast of isolated attosecond pulses is still over $91 \%$. When the incident angle continues to decrease, the corresponding isolated attosecond pulse contrast will also decrease.
\begin{figure}[h]
    \centering
    \includegraphics[width=0.8\textwidth]{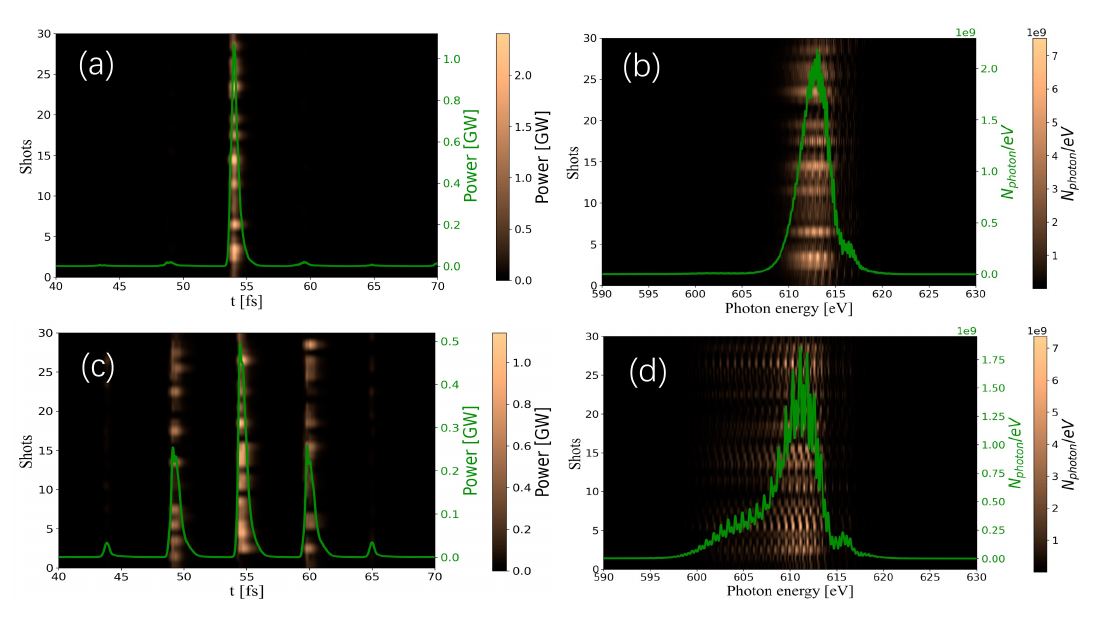}
    \caption{Multi-shot FEL pulses and corresponding spectra in the proposed scenario (a), (b), as well as in the standard ESASE scenario (c), (d). (Green lines represent multi-shot averages. All simulations include LSC effect and taper compensation.) }
    \label{figure6}
\end{figure}

\begin{figure}[h]
    \centering
    \includegraphics[width=1\textwidth]{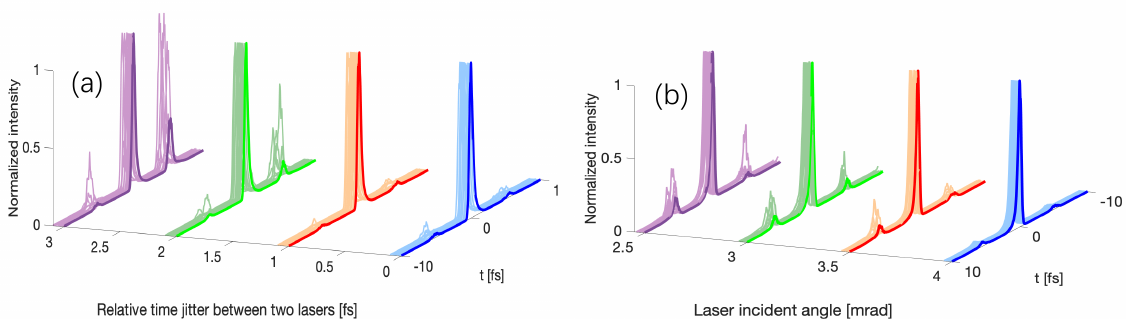}
    \caption{(a) Attosecond FEL performances under different relative time jitters between the two external lasers. (Each dark curve represents the average of ten shots.) (b) Attosecond FEL performances under different laser incident angles. (Each dark curve represents the average of ten shots.) The laser incident angle is  $4 ~\mathrm{mrad}$ in the basic operation mode.}
    \label{figure7}
\end{figure}

\section{Conclusion and discussion}
\label{sec4}  

In this paper, we propose a novel laser manipulation technique for generating isolated attosecond soft X-ray free-electron laser pulses. The novel approach uses two many-cycle laser beams with mutual delays and tilted wavefronts to synthesize microbunching rotation at the infrared laser wavelength. This ultimately leads to an enhanced current contrast ratio between the main peak and the surrounding satellite peaks. By properly accounting for the longitudinal space charge fields within the FEL undulator, a tapered undulator can further suppress the side peaks in the radiation pulse and enable the selection of an isolated GW-level attosecond X-ray pulse. In the standard ESASE scheme, an external laser close to one cycle or even a sub-cycle is usually required to generate isolated attosecond pulses. In the proposed scheme, however, we demonstrate the feasibility of employing many-cycle infrared laser pulses, rather than near-single-cycle lasers, to generate isolated attosecond soft X-ray pulses.
The synthesis approach relies on a three-dimensional laser-beam modulation technique to filter isolated attosecond pulses that will fit down to the current state-of-the-art laser technology. Results demonstrate that the proposed scheme has advantages in generating high-power, high-contrast isolated attosecond soft X-ray pulses. These isolated attosecond pulses with high photon energy and high peak power are of great significance in single-shot imaging in the soft X-ray region, life sciences, and attosecond pump-attosecond probe experiments.

\section*{Acknowledgments}
The authors would like to thank Xiaozhe Shen (IASF), Chao Feng (SARI), Yujie Lu (Zhangjiang Laboratory) and Yaozong Xiao (Zhangjiang Laboratory) for their useful discussions and comments on this work. The authors also would like to express their gratitude to Lingjun Tu (IASF) , Huaiqian Yi (IASF), Yifan Liang (IASF), and Yong Yu (IASF) for their fruitful discussions in physics and simulations. This work is supported by the Scientific Instrument Developing Project of Chinese Academy of Sciences (Grant No. GJJSTD20220001) and the National Natural Science Foundation of China (Grant No. 22288201 and Grant No. 12305359).

\subsection*{Author Contributions} 
Xiaofan Wang and Hao Sun conceived the idea and scheme. Weiqing Zhang and Xiaofan Wang supervised the project. Hao Sun performed the simulations and analyzed the results. Li Zeng engaged in the verification and validation of the simulation results. The contributions of all authors to the writing of the manuscript are equal.

\printbibliography

@ARTICLE{McNeil2010,
   author       = "McNeil, Brian W. J. and Thompson, Neil R.",
   title        = "X-ray free-electron lasers",
   journal      = "Nat. Photonics",
   volume       = "4", 
   pages        = "814-821",
   year         = "2010",
   doi = {10.1038/nphoton.2010.239},
   url = {https://doi.org/10.1038/nphoton.2010.239}
}

@ARTICLE{attosecond1,
   author       = "Sansone, G. and  Kelkensberg, F. and Pérez-Torres, J. F. and Morales, F. and Kling, M. F. and Siu, W. and Ghafur, O. and Johnsson, P. and Swoboda, M. and Benedetti, E. and Ferrari, F. and Lépine, F. and Sanz-Vicario, J. L. and Zherebtsov, S. and Znakovskaya, I. and  L’Huillier, A. and  Ivanov, M. Yu. and Nisoli, M. and Martín, F. and Vrakking, M. J. J.",
   title        = "Electron localization following attosecond molecular photoionization",
   journal      = "Nature",
   volume       = "465", 
   pages        = "763-766",
   year         = "2001",
   doi = {10.1038/35107000},
   url = {https://doi.org/10.1038/35107000}
}

@article{
attosecond2,
author = {F. Calegari  and D. Ayuso  and A. Trabattoni  and L. Belshaw  and S. De Camillis  and S. Anumula  and F. Frassetto  and L. Poletto  and A. Palacios  and P. Decleva  and J. B. Greenwood  and F. Martín  and M. Nisoli },
title = {Ultrafast electron dynamics in phenylalanine initiated by attosecond pulses},
journal = {Science},
volume = {346},
number = {6207},
pages = {336-339},
year = {2014},
doi = {10.1126/science.1254061},
URL = {https://www.science.org/doi/abs/10.1126/science.1254061},
}

@ARTICLE{HHG1,
   author       = "Hentschel, M. and others",
   title        = "Attosecond metrology",
   journal      = "Nature",
   volume       = "414", 
   pages        = "509-513",
   year         = "2001",
   doi = {10.1038/35107000},
   url = {https://doi.org/10.1038/35107000}
}

@ARTICLE{HHG2,
   author       = "Ferrari, F. and others",
   title        = "High-energy isolated attosecond pulses generated by above-saturation few-cycle fields",
   journal      = "Nat. Photonics",
   volume       = "4", 
   pages        = "875-879",
   year         = "2010",
   doi = {10.1038/nphoton.2010.250},
   url = {https://doi.org/10.1038/nphoton.2010.250}
}

@ARTICLE{HHG3,
   author       = "Ossiander, M. and others",
   title        = "Attosecond correlation dynamics",
   journal      = "Nat. Phys.",
   volume       = "13", 
   pages        = "280-285",
   year         = "2017",
   doi = {10.1038/nphys3941},
   url = {https://doi.org/10.1038/nphys3941}
}

@ARTICLE{HHG4,
   author       = "Li, J., Ren and others",
   title        = "53-attosecond X-ray pulses reach the carbon K-edge",
   journal      = "Nat. Commun.",
   volume       = "8", 
   pages        = "186",
   year         = "2017",
   doi = {10.1038/s41467-017-00321-0},
   url = {https://doi.org/10.1038/s41467-017-00321-0}
}

@article{HHG5,
author = {Bernd Sch\"{u}tte and Paul Weber and Katalin Kov\'{a}cs and Emeric Balogh and Bal\'{a}zs Major and Valer Tosa and Songhee Han and Marc J J Vrakking and Katalin Varj\'{u} and Arnaud Rouz\'{e}e},
journal = {Opt. Express},

number = {26},
pages = {33947--33955},
publisher = {Optica Publishing Group},
title = {Bright attosecond soft X-ray pulse trains by transient phase-matching in two-color high-order harmonic generation},
volume = {23},
month = {Dec},
year = {2015},
url = {https://opg.optica.org/oe/abstract.cfm?URI=oe-23-26-33947},
doi = {10.1364/OE.23.033947},

}

@ARTICLE{attoseond_light_house,
   author       = "Wheeler, Jonathan A. and  Borot, Antonin and Monchocé, Sylvain and Vincenti, Henri and Ricci, Aurélien and Malvache, Arnaud and  Lopez-Martens, Rodrigo and Quéré, Fabien",
   title        = "Attosecond lighthouses from plasma mirrors",
   journal      = "Nat. Photonics",
   volume       = "6", 
   pages        = "829",
   year         = "2012",
   doi = {10.1038/nphoton.2012.284},
   url = {https://doi.org/10.1038/nphoton.2012.284}
}

@ARTICLE{Attosecond_control,
   author       = "Baltuška, A. and  Udem, Th. and Uiberacker, M. and others",
   title        = "Attosecond control of electronic processes by intense light fields",
   journal      = "Nature",
   volume       = "421", 
   pages        = "611-615",
   year         = "2003",
   doi = {10.1038/nature01414},
   url = {https://doi.org/10.1038/nature01414}
}

@article{
science.1157846,
author = {E. Goulielmakis  and M. Schultze  and M. Hofstetter  and V. S. Yakovlev  and J. Gagnon  and M. Uiberacker  and A. L. Aquila  and E. M. Gullikson  and D. T. Attwood  and R. Kienberger  and F. Krausz  and U. Kleineberg },
title = {Single-Cycle Nonlinear Optics},
journal = {Science},
volume = {320},
number = {5883},
pages = {1614-1617},
year = {2008},
doi = {10.1126/science.1157846},
URL = {https://www.science.org/doi/abs/10.1126/science.1157846},
}

@article{
science.1132838,
author = {G. Sansone  and E. Benedetti  and F. Calegari  and C. Vozzi  and L. Avaldi  and R. Flammini  and L. Poletto  and P. Villoresi  and C. Altucci  and R. Velotta  and S. Stagira  and S. De Silvestri  and M. Nisoli },
title = {Isolated Single-Cycle Attosecond Pulses},
journal = {Science},
volume = {314},
number = {5798},
pages = {443-446},
year = {2006},
doi = {10.1126/science.1132838},
URL = {https://www.science.org/doi/abs/10.1126/science.1132838},
}

@article{PhysRevLett.100.103906,
  title = {Double Optical Gating of High-Order Harmonic Generation with Carrier-Envelope Phase Stabilized Lasers},
  author = {Mashiko, Hiroki and Gilbertson, Steve and Li, Chengquan and Khan, Sabih D. and Shakya, Mahendra M. and Moon, Eric and Chang, Zenghu},
  journal = {Phys. Rev. Lett.},
  volume = {100},
  issue = {10},
  pages = {103906},
  numpages = {4},
  year = {2008},
  month = {Mar},
  publisher = {American Physical Society},
  doi = {10.1103/PhysRevLett.100.103906},
  url = {https://link.aps.org/doi/10.1103/PhysRevLett.100.103906}
}

@article{RevModPhys.86.897,
  title = {Beam by design: Laser manipulation of electrons in modern accelerators},
  author = {Hemsing, Erik and Stupakov, Gennady and Xiang, Dao and Zholents, Alexander},
  journal = {Rev. Mod. Phys.},
  volume = {86},
  issue = {3},
  pages = {897--941},
  numpages = {45},
  year = {2014},
  month = {Jul},
  publisher = {American Physical Society},
  doi = {10.1103/RevModPhys.86.897},
  url = {https://link.aps.org/doi/10.1103/RevModPhys.86.897}
}

@article{chirp_taper1,
  title = {Self-amplified spontaneous emission FEL with energy-chirped electron beam and its application for generation of attosecond x-ray pulses},
  author = {Saldin, E. L. and Schneidmiller, E. A. and Yurkov, M. V.},
  journal = {Phys. Rev. ST Accel. Beams},
  volume = {9},
  issue = {5},
  pages = {050702},
  numpages = {6},
  year = {2006},
  month = {May},
  publisher = {American Physical Society},
  doi = {10.1103/PhysRevSTAB.9.050702},
  url = {https://link.aps.org/doi/10.1103/PhysRevSTAB.9.050702}
}

@article{chirp_taper2,
  title = {Self-Amplified Spontaneous Emission Free-Electron Laser with an Energy-Chirped Electron Beam and Undulator Tapering},
  author = {Giannessi, L. and others},
  journal = {Phys. Rev. Lett.},
  volume = {106},
  issue = {14},
  pages = {144801},
  numpages = {4},
  year = {2011},
  month = {Apr},
  publisher = {American Physical Society},
  doi = {10.1103/PhysRevLett.106.144801},
  url = {https://link.aps.org/doi/10.1103/PhysRevLett.106.144801}
}

@article{ESASE_2,
  title = {Obtaining attosecond x-ray pulses using a self-amplified spontaneous emission free electron laser},
  author = {Zholents, A. A. and Penn, G.},
  journal = {Phys. Rev. ST Accel. Beams},
  volume = {8},
  issue = {5},
  pages = {050704},
  numpages = {7},
  year = {2005},
  month = {May},
  publisher = {American Physical Society},
  doi = {10.1103/PhysRevSTAB.8.050704},
  url = {https://link.aps.org/doi/10.1103/PhysRevSTAB.8.050704}
}

@article{ESASE_3,
  title = {Generation of attosecond x-ray pulses with a multicycle two-color enhanced self-amplified spontaneous emission scheme},
  author = {Ding, Y. and Huang, Z. and Ratner, D. and Bucksbaum, P. and Merdji, H.},
  journal = {Phys. Rev. ST Accel. Beams},
  volume = {12},
  issue = {6},
  pages = {060703},
  numpages = {6},
  year = {2009},
  month = {Jun},
  publisher = {American Physical Society},
  doi = {10.1103/PhysRevSTAB.12.060703},
  url = {https://link.aps.org/doi/10.1103/PhysRevSTAB.12.060703}
}

@ARTICLE{ESASE_5,
   author       = "Duris, Joseph and others",
   title        = "Tunable isolated attosecond X-ray pulses with gigawatt peak power from a free-electron laser",
   journal      = "Nat. Photonics",
   volume       = "14", 
   pages        = "30-36",
   year         = "2020",
   doi = {10.1038/s41566-019-0549-5},
   url = {https://doi.org/10.1038/s41566-019-0549-5}
}

@article{
ESASE_6,
author = {Siqi Li  and others},
title = {Attosecond coherent electron motion in Auger-Meitner decay},
journal = {Science},
volume = {375},
number = {6578},
pages = {285-290},
year = {2022},
doi = {10.1126/science.abj2096},
URL = {https://www.science.org/doi/abs/10.1126/science.abj2096},
}

@ARTICLE{TW,
   author       = "Franz, Paris and Li, Siqi and Driver, Taran and others",
   title        = "Terawatt-scale attosecond X-ray pulses from a cascaded superradiant free-electron laser",
   journal      = "Nat. Photonics",
   SN  = "1749-4893",

   year         = "2024",
   doi = {10.1038/s41566-024-01427-w},
   url = { https://doi.org/10.1038/s41566-024-01427-w}
}

@article{PhysRevSTAB.12.060701,
  title = {Generation of intense attosecond x-ray pulses using ultraviolet laser induced microbunching in electron beams},
  author = {Xiang, D. and Huang, Z. and Stupakov, G.},
  journal = {Phys. Rev. ST Accel. Beams},
  volume = {12},
  issue = {6},
  pages = {060701},
  numpages = {7},
  year = {2009},
  month = {Jun},
  publisher = {American Physical Society},
  doi = {10.1103/PhysRevSTAB.12.060701},
  url = {https://link.aps.org/doi/10.1103/PhysRevSTAB.12.060701}
}

@article{
Xiao,
author = {Yaozong Xiao  and Chao Feng  and Bo Liu },
title = {Generating Isolated Attosecond X-Ray Pulses by Wavefront Control in a Seeded Free-Electron Laser},
journal = {Ultrafast Sci.},
volume = {2022},
number = {},
pages = {},
year = {2022},
doi = {10.34133/2022/9812478},
URL = {https://spj.science.org/doi/abs/10.34133/2022/9812478},

}

@ARTICLE{reiche1999genesis,
   author       = "S. Reiche",
   title        = "GENESIS 1.3: a fully 3D time-dependent FEL simulation code",
   journal      = "Nucl. Instrum. Methods Phys. Res., Sect. A",
   volume       = "429", 
   pages        = "243",
   year         = "1999",
   publisher={Elsevier},
   doi = {https://doi.org/10.1016/S0168-9002(99)00114-X},
   url = {https://www.sciencedirect.com/science/article/pii/S016890029900114X}
}

@article{PhysRevSTAB.17.070701,
  title = {Three-dimensional manipulation of electron beam phase space for seeding soft x-ray free-electron lasers},
  author = {Feng, Chao and Zhang, Tong and Deng, Haixiao and Zhao, Zhentang},
  journal = {Phys. Rev. ST Accel. Beams},
  volume = {17},
  issue = {7},
  pages = {070701},
  numpages = {7},
  year = {2014},
  month = {Jul},
  publisher = {American Physical Society},
  doi = {10.1103/PhysRevSTAB.17.070701},
  url = {https://link.aps.org/doi/10.1103/PhysRevSTAB.17.070701}
}

@article{PhysRevAccelBeams.22.070701,
  title = {Obliquely incident laser and electron beam interaction in an undulator},
  author = {Wang, Xiaofan and Feng, Chao and Tsai, Cheng-Ying and Zeng, Li and Zhao, Zhentang},
  journal = {Phys. Rev. Accel. Beams},
  volume = {22},
  issue = {7},
  pages = {070701},
  numpages = {10},
  year = {2019},
  month = {Jul},
  publisher = {American Physical Society},
  doi = {10.1103/PhysRevAccelBeams.22.070701},
  url = {https://link.aps.org/doi/10.1103/PhysRevAccelBeams.22.070701}
}

@article{LU2022105849,
title = {Methods for enhancing the steady-state microbunching in storage rings},
journal = {Results Phys.},
volume = {40},
pages = {105849},
year = {2022},
issn = {2211-3797},
doi = {https://doi.org/10.1016/j.rinp.2022.105849},
url = {https://www.sciencedirect.com/science/article/pii/S2211379722004958},
author = {Yujie Lu and Xiaofan Wang and Xiujie Deng and Chao Feng and Dong Wang},
}

@article{Wang_2020,
doi = {10.1088/1367-2630/ab8e5d},
url = {https://dx.doi.org/10.1088/1367-2630/ab8e5d},
year = {2020},
month = {jun},
publisher = {IOP Publishing},
volume = {22},
number = {6},
pages = {063034},
author = {Xiaofan Wang and Chao Feng and Chuan Yang and Li Zeng and Zhentang Zhao},
title = {Transverse-to-longitudinal emittance-exchange in optical wavelength},
journal ={New J. Phys.}
}

@article{Heyl_2014,
doi = {10.1088/1367-2630/16/5/052001},
url = {https://dx.doi.org/10.1088/1367-2630/16/5/052001},
year = {2014},
month = {may},
publisher = {IOP Publishing},
volume = {16},
number = {5},
pages = {052001},
author = {C M Heyl and S N Bengtsson and S Carlström and J Mauritsson and C L Arnold and A. L'Huillier},
title = {Noncollinear optical gating},
journal = {New J. Phys.},

}

@article{Zholents_2008,
doi = {10.1088/1367-2630/10/2/025005},
url = {https://dx.doi.org/10.1088/1367-2630/10/2/025005},
year = {2008},
month = {feb},
publisher = {},
volume = {10},
number = {2},
pages = {025005},
author = {A A Zholents and M S Zolotorev},
title = {Attosecond x-ray pulses produced by ultra short transverse slicing via laser electron beam interaction},
journal = {New J. Phys.}
}

@article{Kennedy_2022,
doi = {10.1088/1367-2630/ac9b80},
url = {https://dx.doi.org/10.1088/1367-2630/ac9b80},
year = {2022},
month = {nov},
publisher = {IOP Publishing},
volume = {24},
number = {11},
pages = {113004},
author = {J P Kennedy and B Dromey and M Yeung},
title = {Isolated ultra-bright attosecond pulses via non-collinear gating},
journal = {New J. Phys.},

}

@article{Louisy:15,
author = {M. Louisy and C. L. Arnold and M. Miranda and E. W. Larsen and S. N. Bengtsson and D. Kroon and M. Kotur and D. Gu\'{e}not and L. Rading and P. Rudawski and F. Brizuela and F. Campi and B. Kim and A. Jarnac and A. Houard and J. Mauritsson and P. Johnsson and A. L'Huillier and C. M. Heyl},
journal = {Optica},
number = {6},
pages = {563--566},
publisher = {Optica Publishing Group},
title = {Gating attosecond pulses in a noncollinear geometry},
volume = {2},
month = {Jun},
year = {2015},
url = {https://opg.optica.org/optica/abstract.cfm?URI=optica-2-6-563},
doi = {10.1364/OPTICA.2.000563},

}

@article{
science.adn6059,
author = {Shuai Li  and Lixin Lu  and Swarnendu Bhattacharyya  and Carolyn Pearce  and Kai Li  and Emily T. Nienhuis  and Gilles Doumy  and R. D. Schaller  and S. Moeller  and M.-F. Lin  and G. Dakovski  and D. J. Hoffman  and D. Garratt  and Kirk A. Larsen  and J. D. Koralek  and C. Y. Hampton  and D. Cesar  and Joseph Duris  and Z. Zhang  and Nicholas Sudar  and James P. Cryan  and A. Marinelli  and Xiaosong Li  and Ludger Inhester  and Robin Santra  and Linda Young },
title = {Attosecond-pump attosecond-probe x-ray spectroscopy of liquid water},
journal = {Science},
volume = {383},
number = {6687},
pages = {1118-1122},
year = {2024},
doi = {10.1126/science.adn6059},
URL = {https://www.science.org/doi/abs/10.1126/science.adn6059},

}

@article{PhysRevLett.114.044801,
  title = {Proposal to Generate an Isolated Monocycle X-Ray Pulse by Counteracting the Slippage Effect in Free-Electron Lasers},
  author = {Tanaka, Takashi},
  journal = {Phys. Rev. Lett.},
  volume = {114},
  issue = {4},
  pages = {044801},
  numpages = {5},
  year = {2015},
  month = {Jan},
  publisher = {American Physical Society},
  doi = {10.1103/PhysRevLett.114.044801},
  url = {https://link.aps.org/doi/10.1103/PhysRevLett.114.044801}
}

@article{APL10.1063/5.0067074,
    author = {Shim, Chi Hyun and Nam, Ki Moon and Parc, Yong Woon and Kim, Dong Eon},
    title = "{Isolated terawatt sub-attosecond high-energy x-ray pulse generated by an x-ray free-electron laser}",
    journal = {APL Photonics},
    volume = {7},
    number = {5},
    pages = {056105},
    year = {2022},
    month = {05},
    
    issn = {2378-0967},
    doi = {10.1063/5.0067074},
    url = {https://doi.org/10.1063/5.0067074},
    eprint = {https://pubs.aip.org/aip/app/article-pdf/doi/10.1063/5.0067074/16493372/056105\_1\_online.pdf},
}

@article{
doi:10.1126/science.287.5461.2237,
author = {R. W. Schoenlein  and S. Chattopadhyay  and H. H. W. Chong  and T. E. Glover  and P. A. Heimann  and C. V. Shank  and A. A. Zholents  and M. S. Zolotorev },
title = {Generation of Femtosecond Pulses of Synchrotron Radiation},
journal = {Science},
volume = {287},
number = {5461},
pages = {2237-2240},
year = {2000},
doi = {10.1126/science.287.5461.2237},
URL = {https://www.science.org/doi/abs/10.1126/science.287.5461.2237},
}

@article{
doi:10.1126/science.1055718,
author = {Patrick G. O'Shea  and Henry P. Freund },
title = {Free-Electron Lasers: Status and Applications},
journal = {Science},
volume = {292},
number = {5523},
pages = {1853-1858},
year = {2001},
doi = {10.1126/science.1055718},
URL = {https://www.science.org/doi/abs/10.1126/science.1055718},
eprint = {https://www.science.org/doi/pdf/10.1126/science.1055718}}

@article{Seddon_2017,
doi = {10.1088/1361-6633/aa7cca},
url = {https://dx.doi.org/10.1088/1361-6633/aa7cca},
year = {2017},
month = {oct},
publisher = {IOP Publishing},
volume = {80},
number = {11},
pages = {115901},
author = {E A Seddon and J A Clarke and D J Dunning and C Masciovecchio and C J Milne and F Parmigiani and D Rugg and J C H Spence and N R Thompson and K Ueda and S M Vinko and J S Wark and W Wurth},
title = {Short-wavelength free-electron laser sources and science: a review*},
journal = {Reports on Progress in Physics}
}

@Article{chirp_taper7,
author = {Schneidmiller, Evgeny and Dreimann, Matthias and Kuhlmann, Marion and Rönsch-Schulenburg, Juliane and Zacharias, Helmut},
title = {Generation of Ultrashort Pulses in XUV and X-ray FELs via an Excessive Reverse Undulator Taper},
journal = {Photonics},
volume = {10},
year = {2023},
pages = {653},
URL = {https://www.mdpi.com/2304-6732/10/6/653},
DOI = {10.3390/photonics10060653}
}

@ARTICLE{Nature_atto_seededFEL,
   author       = "Maroju, Praveen Kumar and Grazioli, Cesare and Di Fraia, Michele and Moioli, Matteo and Ertel, Dominik and Ahmadi, Hamed and Plekan, Oksana and Finetti, Paola and Allaria, Enrico and Giannessi, Luca and others",
   title        = "Attosecond pulse shaping using a seeded free-electron laser",
   journal      = "Nature",
   volume       = "578", 
   pages        = "386-391",
   year         = "2020",
   doi = {10.1038/s41586-020-2005-6},
   url = {https://doi.org/10.1038/s41586-020-2005-6}
}

@article{PhysRevLett.131.145001,
  title = {Experimental Demonstration to Control the Pulse Length of Coherent Undulator Radiation by Chirped Microbunching},
  author = {Tanaka, Takashi and Kida, Yuichiro and Hashimoto, Satoshi and Miyamoto, Shuji and Togashi, Tadashi and Tomizawa, Hiromitsu and Gocho, Aoi and Kaneshima, Keisuke and Tanaka, Yoshihito},
  journal = {Phys. Rev. Lett.},
  volume = {131},
  issue = {14},
  pages = {145001},
  numpages = {6},
  year = {2023},
  month = {Oct},
  publisher = {American Physical Society},
  doi = {10.1103/PhysRevLett.131.145001},
  url = {https://link.aps.org/doi/10.1103/PhysRevLett.131.145001}
}

@Article{photonics7040117,
AUTHOR = {Zeng, Li and Feng, Chao and Wang, Xiaofan and Zhang, Kaiqing and Qi, Zheng and Zhao, Zhentang},
TITLE = {A Super-Fast Free-Electron Laser Simulation Code for Online Optimization},
JOURNAL = {Photonics},
VOLUME = {7},
YEAR = {2020},
NUMBER = {4},
ARTICLE-NUMBER = {117},
URL = {https://www.mdpi.com/2304-6732/7/4/117},
ISSN = {2304-6732},
DOI = {10.3390/photonics7040117}
}

@article{PhysRevSTAB.13.020703,
  title = {Measurements of the linac coherent light source laser heater and its impact on the x-ray free-electron laser performance},
  author = {Huang, Z. and Brachmann, A. and Decker, F.-J. and Ding, Y. and Dowell, D. and Emma, P. and Frisch, J. and Gilevich, S. and Hays, G. and Hering, Ph. and Iverson, R. and Loos, H. and Miahnahri, A. and Nuhn, H.-D. and Ratner, D. and Stupakov, G. and Turner, J. and Welch, J. and White, W. and Wu, J. and Xiang, D.},
  journal = {Phys. Rev. ST Accel. Beams},
  volume = {13},
  issue = {2},
  pages = {020703},
  numpages = {12},
  year = {2010},
  month = {Feb},
  publisher = {American Physical Society},
  doi = {10.1103/PhysRevSTAB.13.020703},
  url = {https://link.aps.org/doi/10.1103/PhysRevSTAB.13.020703}
}

@article{Feng:15,
author = {Chao Feng and Dao Xiang and Haixiao Deng and Dazhang Huang and Dong Wang and Zhentang Zhao},
journal = {Opt. Express},
number = {11},
pages = {14993--15002},
publisher = {Optica Publishing Group},
title = {Generating intense fully coherent soft x-ray radiation based on a laser-plasma accelerator},
volume = {23},
month = {Jun},
year = {2015},
url = {https://opg.optica.org/oe/abstract.cfm?URI=oe-23-11-14993},
doi = {10.1364/OE.23.014993},
}

\end{document}